\documentclass[aps,reprint,prl,superscriptaddress,showpacs,floats,floatfix]{revtex4-2}
\usepackage{graphicx}
\usepackage{amsmath,graphicx,dcolumn}
\usepackage[usenames]{color}
\usepackage{datetime}
\usepackage{textcomp}
\usepackage{ulem}
\usepackage{morefloats}
\usepackage{mathptmx}
\usepackage{indentfirst}
\usepackage{mathrsfs}
\usepackage{bm}
\usepackage{siunitx}
\usepackage{ulem}
\usepackage[colorlinks=true, linkcolor=blue, citecolor=blue, urlcolor=blue]{hyperref}

\newcommand{\blue}{\textcolor{blue}}

\begin{document}

\title{Cleavage-History-Dependent Low-Temperature ARPES Spectra of Charge-Ordered EuAl$_4$} 

\author{Hao Liu}
\affiliation{School of Materials Science and Engineering, Central South University, Changsha 410083, Hunan, China}
\affiliation{School of Physics, Central South University, Changsha 410083, Hunan, China}

\author{Bo Chen}
\affiliation{School of Physics, Central South University, Changsha 410083, Hunan, China}

\author{Chen Zhang}
\affiliation{School of Physics, Central South University, Changsha 410083, Hunan, China}

\author{Qi-Yi Wu}
\affiliation{School of Physics, Central South University, Changsha 410083, Hunan, China}

\author{Sheng-Tao Cui}
\affiliation{National Synchrotron Radiation Laboratory, University of Science and Technology of China, Hefei 230029, Anhui, China}

\author{Zhe Sun}
\affiliation{National Synchrotron Radiation Laboratory, University of Science and Technology of China, Hefei 230029, Anhui, China}

\author{Zhong-Tuo Fu}
\affiliation{School of Physics, Central South University, Changsha 410083, Hunan, China}

\author{Ying Zhou}
\affiliation{School of Physics, Central South University, Changsha 410083, Hunan, China}

\author{Yang Luo}
\affiliation{School of Physics, Central South University, Changsha 410083, Hunan, China}

\author{Jun Liu}
\affiliation{School of Materials Science and Engineering, Central South University, Changsha 410083, Hunan, China}

\author{Yu-Xia Duan}
\affiliation{School of Physics, Central South University, Changsha 410083, Hunan, China}

\author{Jian-Qiao Meng}
\email{Corresponding author: jqmeng@csu.edu.cn}\affiliation{School of Physics, Central South University, Changsha 410083, Hunan, China}

\date{\today}

\begin{abstract}

Charge ordering in EuAl$_4$ has been widely discussed in connection with band reconstruction, magnetism, and topological electronic states, yet the microscopic origin of the complex low-temperature ARPES spectra remains unresolved. Here we combine photon-energy-, temperature-, and cleavage-history-dependent ARPES with first-principles calculations to distinguish intrinsic bulk bands from surface-preparation-dependent spectral weight. Spectra measured on high-temperature-cleaved surfaces, both at 160 K and after cooling to 10 K, are broadly consistent with the calculated three-dimensional bulk electronic structure, whereas low-temperature-cleaved surfaces exhibit additional electron-like bands, replica-like Fermi-surface contours, and a pronounced $\delta$ band near -0.57 eV that is absent from the calculated bulk bands. The additional features are observed at multiple photon energies and on multiple independently cleaved surfaces and are selectively suppressed upon warming, while the bulk-derived bands remain comparatively stable. The $\delta$ band does not emerge when the same high-temperature-cleaved surface is cooled through $T_{\rm CDW}$. Comparison with the projected bulk bands and the calculated spectral function of an ideal Eu-terminated surface further associates the additional bands with the surface electronic structure. These results establish a strong cleavage-history dependence of the low-temperature ARPES spectra and provide spectroscopic criteria for separating surface-reconstruction and bulk charge-order contributions in EuAl$_4$.

\end{abstract}

\maketitle

EuAl$_4$ belongs to the BaAl$_4$-family square-net semimetals, in which charge order, local-moment magnetism, and nontrivial band topology coexist within a relatively simple tetragonal structure \cite{JMMoya2023, SLei2023, TShang2024JPCM}. In EuAl$_4$, a charge-density-wave (CDW) transition occurs at $T_{\rm CDW} \approx$ 140 K \cite{KKaneko2021, SShimomura2019}, followed by a sequence of antiferromagnetic and field-induced magnetic phases at lower temperatures \cite{SShimomura2019, WRMeier2022, RTakagi2022, MGen2023, RYang2024PRB, AMVibhakar2024, HMiao2024PRX}. These intertwined ordered states make EuAl$_4$ a model system for studying how itinerant carriers in Al-derived bands couple to Eu local moments, lattice modulations, and topological band crossings. A reliable identification of the intrinsic bulk electronic structure is therefore a prerequisite for understanding the microscopic origin of its CDW instability, Ruderman-Kittel-Kasuya-Yosida (RKKY)-mediated magnetic interactions, and field-induced topological magnetic phases.

Angle-resolved photoemission spectroscopy (ARPES) provides a direct, momentum-resolved probe of the Fermi-surface topology, nesting tendencies, and band renormalization in correlated materials \cite{BQLv2021, BChen2024, CZhang2023}. Previous photoemission and band-structure studies established that EuAl$_4$ and related BaAl$_4$-type compounds host a three-dimensional (3D) multiband electronic structure with multiple Fermi-surface sheets and Dirac-like crossings near the Fermi level \cite{HMiao2024PRX, LWang2024CP, MKobata2016, YArai2026NC, YArai2026PRB, AEaton2024, TLi2026}. In a conventional CDW scenario, ARPES should reveal either partial gap opening at nested Fermi-surface segments or band folding associated with the CDW wave vector \cite{CZhang2022, ZTLiu2023, YMSun2024}. However, low-temperature ARPES spectra of EuAl$_4$ often display substantially richer structures than expected from calculated bulk bands, including enlarged electron-like pockets, replica-like Fermi-surface contours, additional low-energy dispersions, and spectral features showing little photon-energy dependence \cite{YArai2026NC, YArai2026PRB, AEaton2024, TLi2026}.

\begin{figure*}[tbp]
\vspace*{-0.2cm}
\begin{center}
\includegraphics[width=1.98\columnwidth]{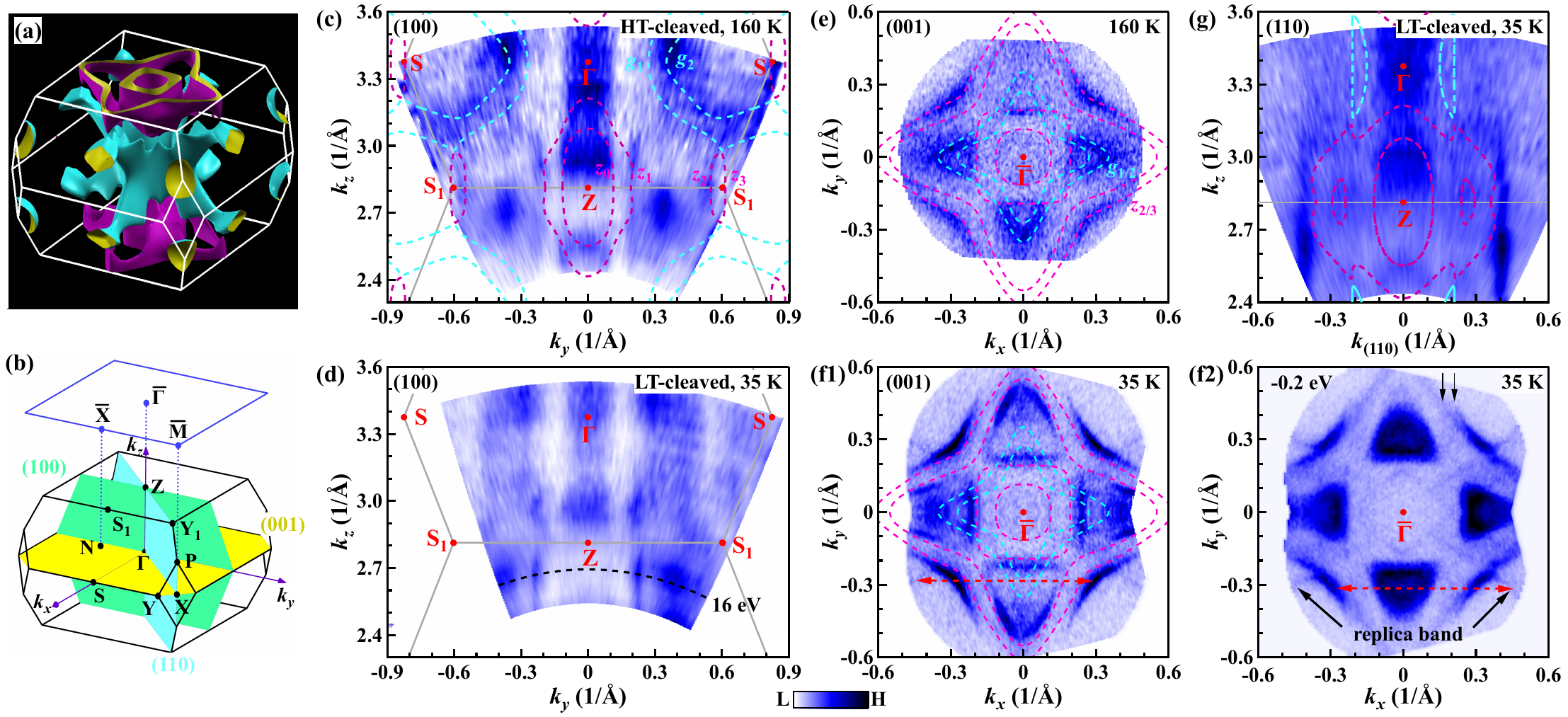}
\end{center}
\vspace*{-0.5cm}
\caption{\label{FIG:1}\textbf{Calculated and measured Fermi surfaces of EuAl$_4$.} (a) DFT-calculated 3D bulk Fermi surface. (b) Bulk Brillouin zone with the ARPES momentum planes indicated. (c), (d) Photon-energy-dependent ARPES intensity maps in the $k_y$-$k_z$ plane measured on HT-cleaved and LT-cleaved surfaces, respectively. Colored dashed curves denote calculated bulk Fermi-surface contours; the black dashed line marks the 16 eV cut. (e) In-plane $k_x$-$k_y$ Fermi-surface maps measured with $h\nu$ = 16 eV at 160 K on an HT-cleaved surface, with calculated $\Gamma$- and Z-plane contours overlaid. (f1), (f2) In-plane constant-energy contours measured at 35 K on an LT-cleaved surface at $E_F$ and $E$ - $E_F$ = -0.2 eV, respectively. (g) Photon-energy-dependent ARPES intensity map in the $k_{(110)}$-$k_z$ plane measured at 35 K. The intense outermost branch, corresponding to the arc-like feature in (f1), shows little photon-energy dependence, in contrast to the bulk-derived contours.}
\end{figure*}

This ambiguity is not merely technical. Because ARPES is surface sensitive, cleavage-induced surface reconstruction can generate electronic states that overlap strongly with bulk bands, especially in layered or quasi-layered correlated materials \cite{VNStrocov2003, SFujimori2016, YZZhao2024}. For EuAl$_4$(001), scanning tunneling microscopy has directly revealed multiple cleavage-dependent surface terminations, half-unit-cell steps, adsorbed Eu atoms, and a vacancy-ordered 2$\times$1 surface reconstruction \cite{TLi2026, JLGrant2026}. Such surface modifications can produce quasi-two-dimensional (quasi-2D) states, in-plane replica bands, and reconstructed Fermi-surface contours that are superposed on the bulk spectrum and may complicate the identification of weaker intrinsic bulk-CDW-induced changes. A closely related precedent is Sr$_2$RuO$_4$, where reconstructed surface domains generate folded bands and additional Fermi-surface sheets that must be separated from the intrinsic bulk electronic structure \cite{ADamascelli2000}. These considerations raise the central question addressed here: which parts of the low-temperature reconstructed ARPES spectra of EuAl$_4$ are intrinsic consequences of bulk charge order, and which parts arise from cleavage-dependent surface reconstruction?

To distinguish bulk and surface contributions, we combine photon-energy-, temperature-, and cleavage-history-dependent ARPES with first-principles calculations. Spectra measured on high-temperature-cleaved surfaces at 160 K and after subsequent cooling are broadly consistent with the calculated bulk electronic structure, whereas low-temperature-cleaved surfaces develop additional electron-like bands, replica-like Fermi-surface contours, and a pronounced $\delta$ band near -0.57 eV. Together, these measurements reveal a strong cleavage-history dependence of the low-temperature ARPES spectra of EuAl$_4$.

Single crystals of EuAl$_4$ were grown by the self-flux method \cite{HMiao2024PRX}. High-quality crystals with shiny natural cleavage surfaces were obtained for ARPES measurements (see Sec. \blue{II} of the Supplemental Material \cite{Supplemental Materials}). ARPES measurements were performed at the BL13U beamline of the National Synchrotron Radiation Laboratory with a Scienta DA30 analyzer. The energy and angular resolutions were better than 15 meV and 0.3\textdegree, respectively. The photon footprint on the sample was approximately 0.3 $\times$ 0.3 $mm^2$. Samples were cleaved in situ under a vacuum better than $6 \times 10^{-11}$ mbar. Subsequently, an inner potential ($V_0$) of 16 eV was taken, consistent with findings from other studies \cite{YArai2026NC, YArai2026PRB}.

To separate intrinsic bulk electronic reconstruction from cleavage-induced surface effects, we compared two surface-preparation protocols. High-temperature-cleaved surfaces, denoted as HT-cleaved, were cleaved at 160 K and subsequently cooled when needed. Low-temperature-cleaved surfaces, denoted as LT-cleaved, were cleaved at low temperature, typically 10-35 K, and measured during subsequent warming. Photon-energy-dependent ARPES was used to assess the $k_z$ dependence of the measured spectral features, while temperature-dependent measurements on the same cleaved surface tracked the evolution of the reconstructed spectral weight. Bulk band dispersions and Fermi surfaces were calculated using density-functional theory and Wannier-based tight-binding models including spin-orbit coupling \cite{PEBlochl1994, GKresse1996, JPPerdew1996, NMarzari1997PRB, ISouza2001PRB, QWu2018}. In the bulk reference calculations, the localized Eu 4$f$ electrons were treated as core states and excluded from the low-energy valence manifold. Further computational details are provided in Sec. \blue{I} of the Supplemental Material \cite{Supplemental Materials}.

To establish the bulk electronic reference against which reconstructed low-temperature spectral weight can be identified, we first compare the measured Fermi surfaces with the calculated 3D bulk Fermi surface of EuAl$_4$. The calculated Fermi surface [Fig. \blue{1(a)}] contains multiple electron- and hole-like sheets with pronounced $k_z$ dispersion, reflecting the 3D multiband electronic structure of EuAl$_4$. The relevant bulk Brillouin zone and the measured momentum planes are shown in Fig. \blue{1(b)}. This comparison is essential because a bulk CDW reconstruction should modify the bulk Fermi surface in a manner constrained by the CDW wave vector and should remain reproducible for different cleaved surfaces, whereas a surface reconstruction can generate quasi-2D replica features that do not follow the calculated 3D bulk topology.

The photon-energy-dependent map in the $k_y$-$k_z$ plane, corresponding to the (100) plane, provides the first separation between these two possibilities. For the HT-cleaved surface measured at 160 K, above the CDW transition, the Fermi-surface contours [Fig. \blue{1(c)}] follow the calculated bulk sheets. In particular, the $g_1$ and $g_2$ pockets around $\Gamma$ show pronounced photon-energy dependence, establishing their 3D bulk character. This agreement defines the nonreconstructed bulk reference for the subsequent low-temperature analysis.

On the LT-cleaved surface measured at 35 K, the principal bulk-derived contours remain visible in the same $k_y$-$k_z$ plane, accompanied by a redistribution of spectral weight in their vicinity [Fig. \blue{1(d)}]. The nesting/CDW vector proposed for EuAl$_4$ is small and primarily out of plane, $q_{\rm CDW} \approx$ 0.1 $Å^{-1}$ \cite{SShimomura2019, KKaneko2021}, so a subtle bulk folding signal, if present, would be difficult to isolate in low-photon-energy ARPES because of finite $k_z$ resolution \cite{VNStrocov2003, SFujimori2016} and matrix-element sensitivity. The individual low-temperature spectral components are resolved more clearly in the in-plane Fermi-surface maps and momentum-resolved band dispersions discussed below.

The in-plane Fermi-surface maps at $h\nu$ = 16 eV reveal the momentum-space structure of these additional components. For the HT-cleaved surface measured at 160 K [Fig. \blue{1(e)}], the $k_x$-$k_y$ map shows the main trends of the calculated $\Gamma$- and Z-plane contours. Because the 16 eV spectra sample a broadened $k_z$ window, the overlaid $\Gamma$- and Z-plane calculations are used only as qualitative high-symmetry references. The outer contours are broadly consistent with the Z-plane calculation, whereas the inner pockets associated with the $g_1$ and $g_2$ bands show similarities to the $\Gamma$-plane contribution. This qualitative correspondence supports the use of the high-temperature spectrum as a bulk-dominated reference.

By contrast, the 35 K map obtained on an LT-cleaved surface [Fig. \blue{1(f1)}] contains several features absent from the bulk calculation. The inner triangular pockets are enlarged, an additional smaller pocket appears inside each triangle along $k_y$ = 0, and the outer arc-like contours sharpen into replica-like structures along the direction marked by the dashed arrow. Similar replica-like features have been reported in previous low-temperature ARPES studies on cleaved EuAl$_4$(001) surfaces \cite{TLi2026, YArai2026PRB}. The constant-energy contour at $E$-$E_F$ = -0.2 eV [Fig. \blue{1(f2)}] shows that the triangular pockets shrink with increasing binding energy, identifying them as electron-like bands. At the same time, the outer arcs split into multiple components and connect to a weak diamond-shaped intensity pattern near the zone center, indicating that the low-temperature Fermi surface is not a single reconstructed bulk contour but a superposition of several dispersive components. Recent micro-ARPES studies resolved termination- and domain-dependent spectra on spatial scales of approximately 10-15 $\mu$m \cite{YArai2026PRB, TLi2026}. By comparison, the substantially larger photon footprint used here averages the photoemission signal over an extended surface area. The persistence of the additional bands under this spatial averaging indicates that they contribute appreciable spectral weight to the macroscopic ARPES response.

\begin{figure}[tbp]
\vspace*{-0.2cm}
\begin{center}
\includegraphics[width=0.98 \columnwidth,angle=0]{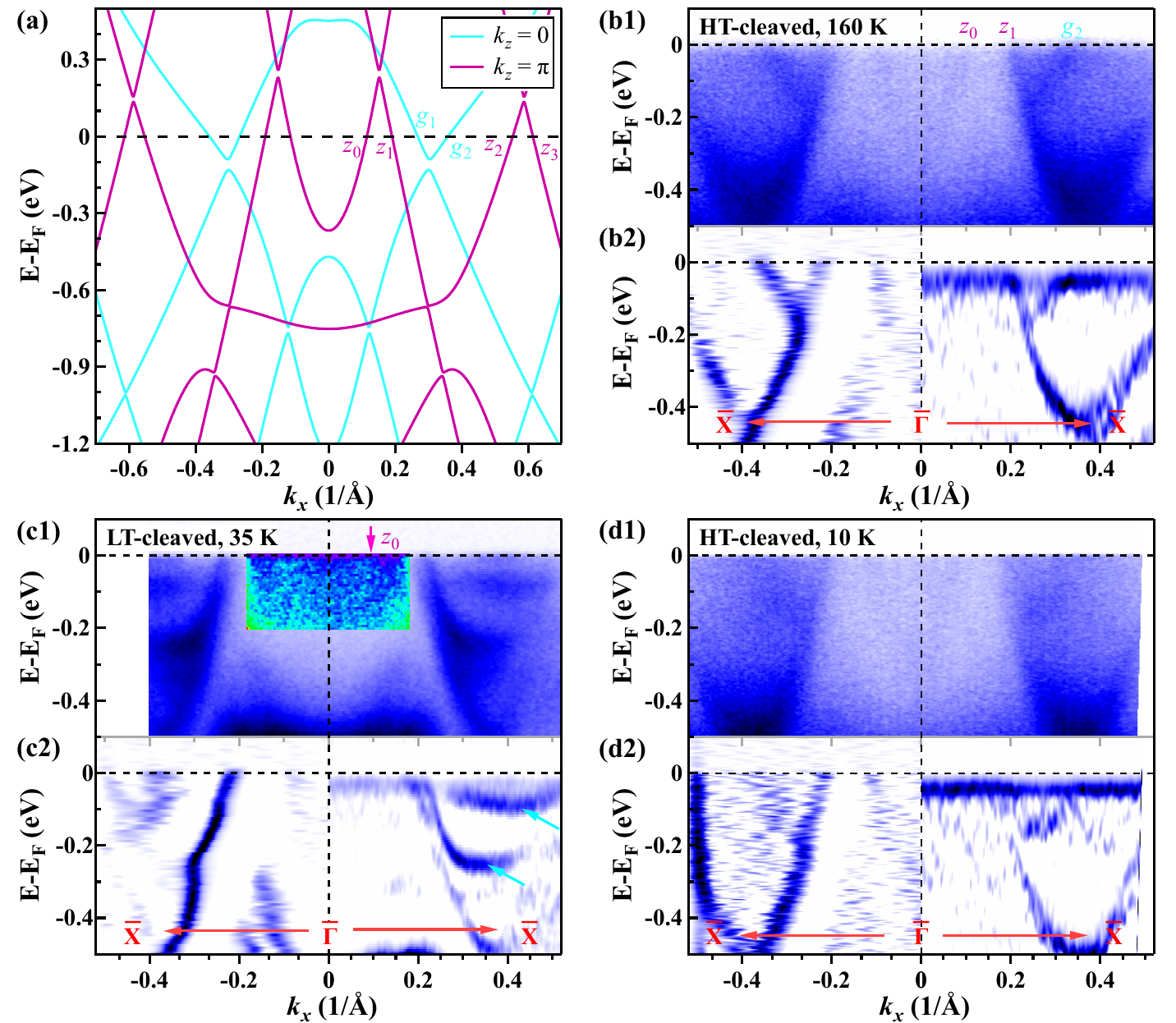}
\end{center}
\vspace*{-0.5cm}
\caption{\label{FIG:2}\textbf{Calculated and measured band dispersions of EuAl$_4$ at $h\nu$ = 16 eV.} (a) DFT-calculated bulk band dispersions along $\rm S$-$\Gamma$-$\rm S$ ($k_z=0$) and $\rm S_1$-Z-$\rm S_1$ ($k_z=\pi$), with near-$E_F$ bands labeled. (b1), (b2) ARPES intensity plot and corresponding second-derivative images measured at 160 K on an HT-cleaved surface. (c1), (c2) Corresponding spectra measured at 35 K on an LT-cleaved surface. Arrows mark two additional near-$E_F$ electron-like bands near the $z_1$ and $z_2$ bulk bands. (d1), (d2) Spectra measured at 10 K after cooling the same HT-cleaved surface used in (b1). For (b2)-(d2), the left and right panels show the momentum- and energy-derivative images, respectively.}
\end{figure}

Further support comes from the photon-energy-dependent measurements in the $k_{\rm (110)}$-$k_z$ plane [Fig. \blue{1(g)}]. Most spectral features disperse strongly with photon energy and follow the calculated bulk Fermi-surface contours, as expected for 3D bulk bands. In contrast, the outermost branch varies little with photon energy. This branch corresponds to the arc-like feature in the in-plane maps, and its weak photon-energy dependence is consistent with a quasi-2D component. The supplemental photon-energy-dependent measurements over 11-23 eV show the same behavior: several prominent features remain nearly unchanged with photon energy while the bulk-related spectral weight evolves with $k_z$ (see Fig. \blue{S2} in Sec. \blue{III} of the Supplemental Material \cite{Supplemental Materials}). Together with the cleavage-history controls and the calculations below, these observations distinguish the calculated 3D bulk Fermi surface from an additional quasi-2D low-temperature component.

To identify the band origin of the additional low-temperature Fermi-surface features, we next examine the corresponding dispersions along the high-symmetry cut shown in Fig. \blue{2}. Figure \blue{2(a)} shows the calculated bulk bands along the $\rm S$-$\Gamma$-$\rm S$ ($k_z = 0$) and $\rm S_1$-Z-$\rm S_1$ ($k_z = \pi$) directions, with the relevant low-energy bands labeled as $z_0$-$z_3$ near the Z plane and $g_1$, $g_2$ near the $\Gamma$ plane. These two calculated cuts provide qualitative high-symmetry references for the 16 eV spectra, which sample a broadened $k_z$ window rather than a single ideal $k_z$ plane.

At 160 K on an HT-cleaved surface [Fig. \blue{2(b1)}], the measured dispersions show the main trends of the calculated bulk bands. Because low-photon-energy ARPES has pronounced $k_z$ broadening, the spectra contain overlapping contributions from both the $\Gamma$- and Z-plane electronic states. Within this broadened $k_z$ window, the dominant observed branches can be associated with the bulk-derived $z_0$, $z_1$, and $g_2$ bands, consistent with the mixed $\Gamma$- and Z-plane Fermi-surface contributions identified in Fig. \blue{1(e)}. This qualitative correspondence supports the use of the high-temperature spectrum as a bulk-dominated reference.

The spectrum measured at 35 K on an LT-cleaved surface [Fig. \blue{2(c1)}] is qualitatively different. In addition to the suppression or masking of the $g_2$ branch, two additional near-$E_F$ electron-like bands appear between the $z_1$ and $z_2$ bulk bands, as marked by cyan arrows in Fig. \blue{2(c2)}. Their band minima are located near -0.25 and -0.08 eV, respectively. The deeper branch accounts for the enlarged triangular pockets observed in Fig. \blue{1(f1)}, whereas the shallower branch gives rise to the smaller inner pocket along $k_y$ = 0. These bands occupy an energy-momentum region where the bulk calculation does not predict corresponding states. Their appearance is therefore not accounted for by a simple overlap of $\Gamma$- and Z-plane bulk bands caused by $k_z$ broadening.

The same HT-cleaved surface was measured at 160 K and again after cooling to 10 K, providing a same-surface cooling control [Fig. \blue{2(d1)}]. As shown in Figs. \blue{2(d1)} and \blue{2(d2)}, the resulting low-temperature spectrum remains close to the 160 K bulk-like dispersion and does not develop the two additional electron-like bands observed on the LT-cleaved surface. Thus, cooling through $T_{\rm CDW}$ alone does not generate the additional bands; their appearance depends strongly on the surface-preparation history.

The cleavage-history comparison identifies the anomalous low-temperature bands as a surface-preparation-dependent component superposed on the bulk spectrum. Their weak photon-energy dependence and replica-like momentum structure in Fig. \blue{1} further connect them with cleavage-dependent surface reconstruction. The broader photon-energy-dependent data in Fig. \blue{S5} of the Supplemental Material \cite{Supplemental Materials} reproduce this behavior, including the appearance of the $\delta$ band near -0.57 eV on LT-cleaved surfaces and its absence on the same HT-cleaved surface after cooling to 10 K.

 \begin{figure}[tbp]
\vspace*{-0.2cm}
\begin{center}
\includegraphics[width=0.98 \columnwidth,angle=0]{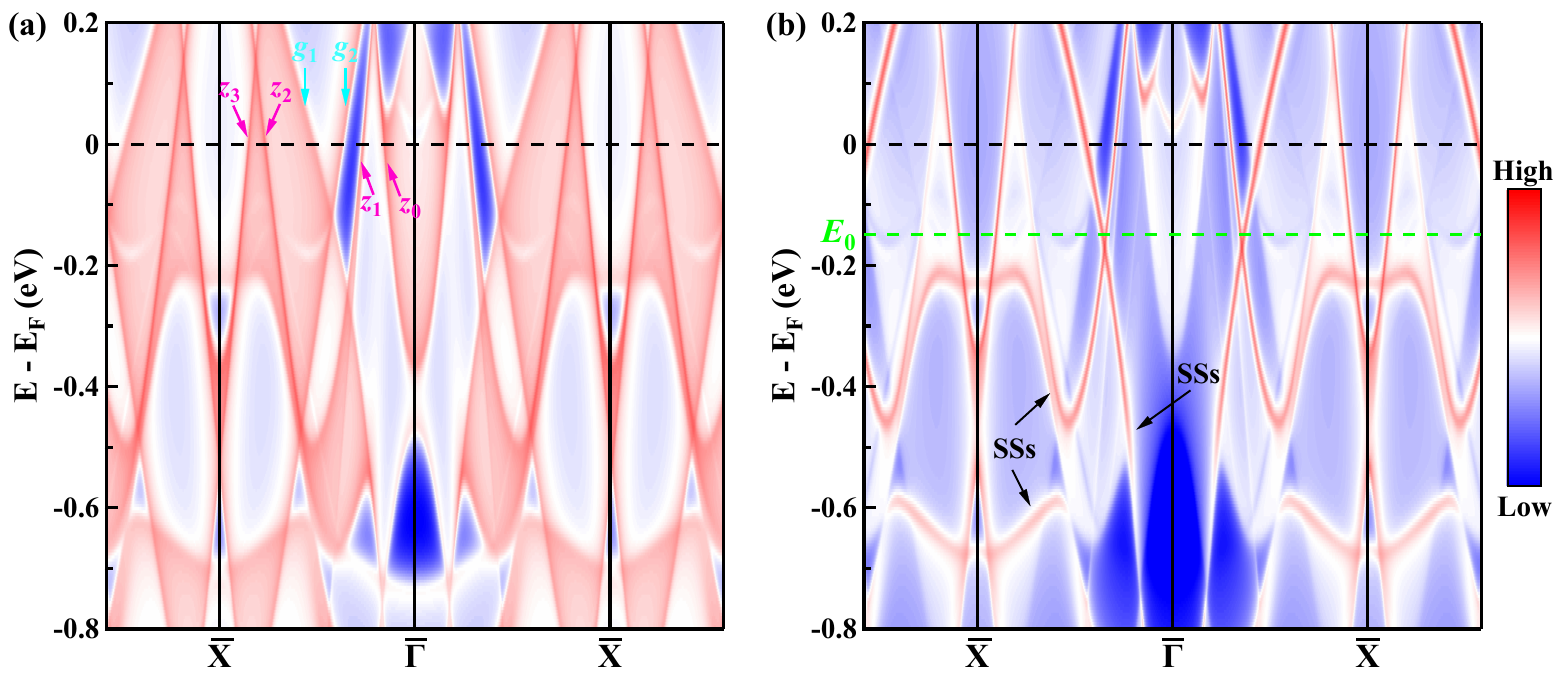}
\end{center}
\vspace*{-0.5cm}
\caption{\label{FIG:2}\textbf{Theoretical bulk projection and surface spectral function of EuAl$_4$ (001).} (a) Bulk-band projection onto the (001) surface Brillouin zone. The projected $\Gamma$-derived $g_1$ and $g_2$ bands and Z-derived $z_0$–$z_3$ bands are labeled. (b) Calculated surface spectral function for the ideal Eu-terminated (001) surface. The green dashed line marks the alignment energy $E_0$ (-0.15 eV). Arrows indicate additional surface-derived states that are absent from the bulk projection.}
\end{figure}

To further clarify the origin of the observed electronic states, we compare the ARPES results with the bulk-band projection onto the (001) surface and the calculated spectral function of an ideal Eu-terminated surface. As shown in Fig. \blue{3(a)}, the projected bulk bands contain both the $\Gamma$-derived $g_1$ and $g_2$ bands and the Z-derived $z_0$–$z_3$ bands, providing a more appropriate reference for spectra with finite $k_z$ resolution than a single-$k_z$ calculation. In contrast, the Eu-terminated surface spectral function [Fig. \blue{3(b)}] contains additional electron-like surface-derived states, indicated by arrows, that are absent from the bulk projection. The green dashed line marks the alignment energy $E_0$. After this energy alignment, the additional calculated branches show an overall energy-momentum correspondence with the electron-like bands observed on LT-cleaved surfaces. The calculation represents an ideal unreconstructed Eu-terminated surface, whereas the cleaved surface can host the reported vacancy-ordered 2$\times$1 reconstruction and multiple terminations \cite{TLi2026, JLGrant2026}; these structural differences may contribute to the remaining dispersion and intensity differences. The correspondence complements the cleavage-history dependence and supports the association of the observed additional bands with the surface electronic structure.

We next examine the temperature evolution of the additional spectral weight on an LT-cleaved surface and compare it with the bulk-derived bands. This measurement tracks the relative stability of the two components over the same warming sequence.

\begin{figure*}[tbp]
\vspace*{-0.2cm}
\begin{center}
\includegraphics[width=1.98\columnwidth]{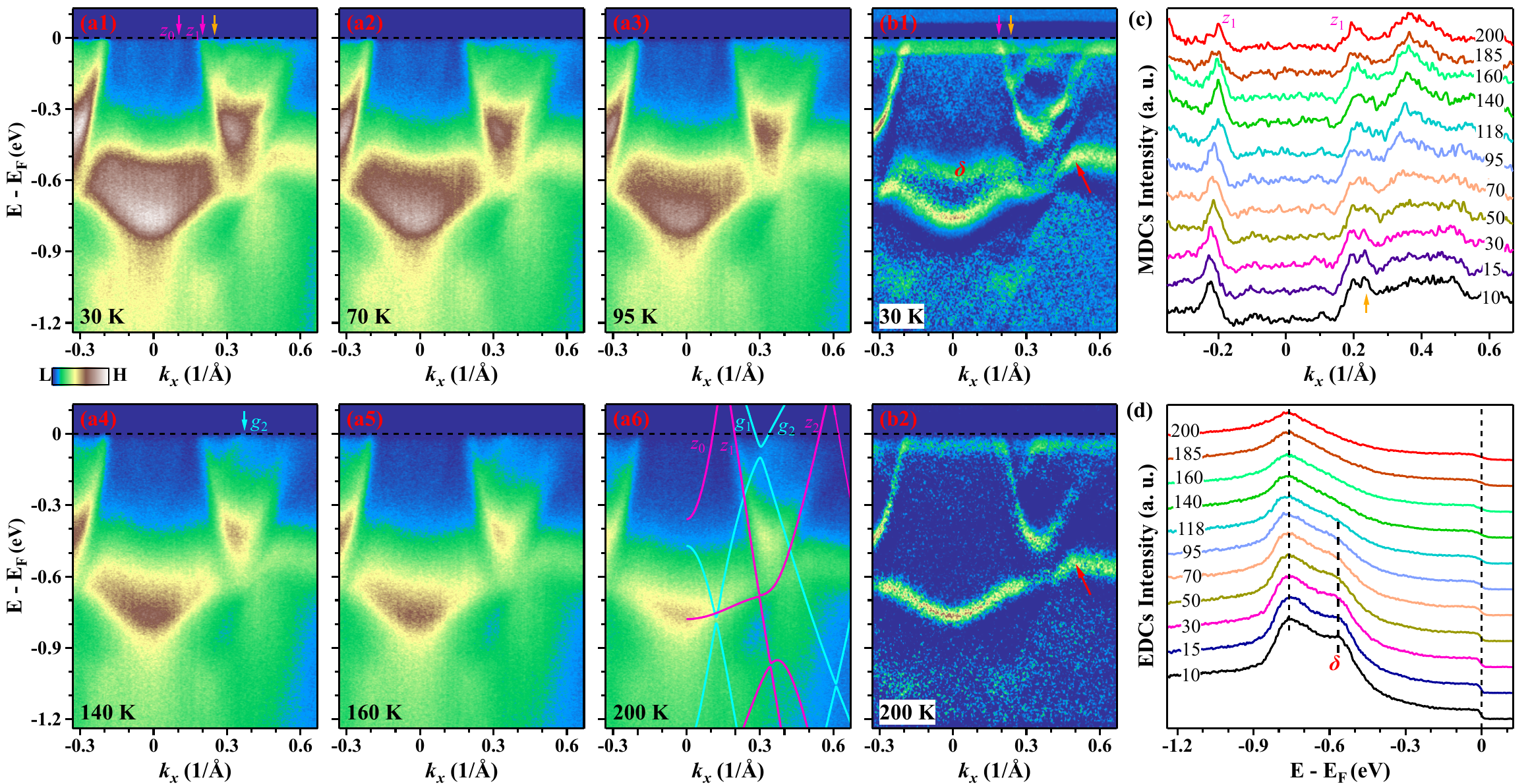}
\end{center}
\vspace*{-0.5cm}
\caption{\label{FIG:3}\textbf{Temperature evolution of an LT-cleaved EuAl$_4$(001) surface at $h\nu$ = 16 eV.} (a1)-(a6) ARPES spectra measured at selected temperatures. (b1), (b2) Second-derivative images taken along the energy direction for the spectra measured at 30 and 200 K, respectively. The shallow electron-like $\delta$ band near -0.57 eV is marked in (b1). (c) Momentum-distribution curves at $E_F$ for selected temperatures. (d) Energy-distribution curves integrated around $k_x$ = 0. The $\delta$-band peak is progressively suppressed upon warming and becomes difficult to distinguish from the background by approximately 118 K.}
\end{figure*}

Figure \blue{4(a)} shows temperature-dependent ARPES spectra measured with $h\nu$ = 16 eV on the same LT-cleaved EuAl$_4$(001) surface. At low temperature, the spectra contain the additional near-$E_F$ electron-like bands identified in Fig. \blue{2}, together with two extra features at higher binding energy [Fig. \blue{4(b1)}]. The first is a shallow electron-like band centered near $k_x$ = 0, with a band bottom around -0.57 eV; we denote this feature as the $\delta$ band. The second appears at larger momentum, $k_x \gtrsim 0.4$ $\mathrm{\AA^{-1}}$, as a wing-like branch. Both features are absent from the calculated bulk bands, indicating that the LT-cleaved surface differs from the intrinsic bulk reference over an extended energy range, not only near $E_F$.

Upon warming, the low-temperature spectral complexity is progressively suppressed. By 200 K [Fig. \blue{4(a6)}], the measured dispersion approaches a simpler bulk-like band structure broadly consistent with the calculated $S$-$\Gamma$-$S$ and $\rm S_1$-Z-$\rm S_1$ dispersions. The dominant branches can be associated with the $z_0$, $z_1$, $z_2$, and $g_2$ bulk bands. The second-derivative images in Figs. \blue{4(b1)} and \blue{4(b2)} make this evolution more apparent: the dense low-temperature spectral weight between the bulk branches is strongly suppressed upon warming, leaving a simpler dispersion consistent with the bulk calculation.

The momentum-distribution curves at $E_F$ provide a quantitative view of this process [Fig. \blue{4(c)}]. The peaks associated with the $z_1$ bulk band, located near $k_x \approx \pm 0.2$ $\mathrm{\AA^{-1}}$, show little momentum shift over the measured temperature range. This stability indicates that the corresponding bulk-derived bands are not undergoing a large temperature-driven reconstruction within our experimental resolution. In contrast, the extra peaks inside the enlarged electron pocket, near $k_x \approx \pm 0.25$ $\mathrm{\AA^{-1}}$, lose spectral weight rapidly upon warming, indicating that this near-$E_F$ component is substantially more temperature sensitive than the bulk-derived bands. The $g_2$-related bulk feature near $k_x \approx 0.35$ $\mathrm{\AA^{-1}}$ is obscured at low temperature but becomes more visible as the overlapping surface-derived spectral weight is suppressed.

The energy-distribution curves integrated around $k_x$ = 0 show the same separation between stable bulk states and temperature-sensitive additional spectral weight [Fig. \blue{4(d)}]. The bulk-derived peak near -0.76 eV remains essentially fixed in energy, whereas the $\delta$-band peak around -0.57 eV weakens rapidly and becomes difficult to distinguish from the background by approximately 118 K. The warming series therefore reveals a selective loss of the additional low-temperature spectral weight while the parent bulk-band structure remains comparatively stable. The contrasting temperature dependences distinguish the additional component from the bulk-derived bands and support its association with the reconstructed surface electronic structure.

The wing-like branch behaves differently from this temperature-sensitive component. The wing-like branch at $k_x \gtrsim 0.4$ $\mathrm{\AA^{-1}}$ persists throughout the measured temperature range and shows no obvious change in dispersion, as indicated in Figs. \blue{4(b1)} and \blue{4(b2)}. Its thermal robustness suggests a microscopic origin distinct from the temperature-sensitive near-$E_F$ pockets and the $\delta$ band, which are suppressed upon warming over different apparent temperature scales. Similar persistent features have been reported in EuGa$_2$Al$_2$ \cite{YArai2026PRB} and BaAl$_4$ \cite{RMori2022}, suggesting that they may represent a more robust surface-related band common to BaAl$_4$-family compounds. Together with the photon-energy and cleavage-history dependences, the selective suppression of the latter features supports the assignment of the dominant additional low-temperature spectral weight to a temperature-sensitive surface-reconstructed component.

In conclusion, photon-energy-, temperature-, and cleavage-history-dependent ARPES measurements, benchmarked against the projected bulk bands and the calculated spectral function of an ideal Eu-terminated surface, separate the intrinsic 3D electronic structure of EuAl$_4$ from surface-derived spectral features. The spectra measured on HT-cleaved surfaces are broadly consistent with the calculated bulk bands, both at 160 K and after cooling to 10 K. In contrast, LT-cleaved surfaces develop additional near-$E_F$ electron-like pockets, replica-like Fermi-surface contours, and a temperature-sensitive $\delta$ band near -0.57 eV, which are absent from the bulk calculation and are selectively suppressed upon warming. The absence of the $\delta$ band after cooling the same HT-cleaved surface through $T_{\rm CDW}$, together with its photon-energy and temperature dependences, supports assigning the dominant additional low-temperature spectral weight to a surface-preparation-dependent electronic component. This distinction provides spectroscopic criteria for separating surface reconstruction from bulk charge-order effects in BaAl$_4$-family correlated semimetals.

This work was supported by the National Key Research and Development Program of China (Grant No. 2022YFA1604204), the Beijing National Laboratory for Condensed Matter Physics (Grant No. 2024BNLCMPKF001), the National Natural Science Foundation of China (Grant No. 12574168), the Science and Technology Innovation
Program of Hunan Province (Grant No. 2022RC3068), and the China Postdoctoral Science Foundation (Grant No. 2025M783459). Part of this work was carried out using the computing resources at the High Performance Computing Center of Central South University.

\textit{Data availability}—The data that support the findings of this article are not publicly available. The data are available from the authors upon reasonable request.

\end{document}